\documentclass{ws-ijmpb}

\begin{document}

\markboth{A. Kregar and A. Ram{\v s}ak}
{Exact unitary transformation for Rashba rings in magnetic and electric fields}

%
\catchline{}{}{}{}{}
%

\title{EXACT UNITARY TRANSFORMATION FOR RASHBA RINGS \\ IN MAGNETIC AND ELECTRIC FIELDS}

\author{A.  Kregar}

\address{J. Stefan Institute, Ljubljana, Slovenia\\
Faculty of Mechani­cal Engineering, University of Ljubljana, Ljubljana, Slovenia\\
ambroz.kregar@ijs.si}

\author{A. Ram\v{s}ak}

\address{J. Stefan Institute, Ljubljana, Slovenia\\
Faculty of Mathematics and Physics, University of Ljubljana, Ljubljana, Slovenia\\
anton.ramsak@fmf.uni-lj.si}

\maketitle

\begin{history}
\end{history}

\begin{abstract}
An exact solution for single electron states on mezoscopic rings with the Rashba coupling and in the presence of external magnetic and electric fields is derived by means of a unitary transformation. The transformation maps the model to a bare ring, which gives the possibility of a very simple formulation of single or many electron problems. As an example some exact results for spin and energy levels are presented.
\end{abstract}

\keywords{Rashba interaction; ring system; unitary transformation.}

\section{Introduction}
In classical electronics the electron charge is a fundamental resource, while in the new branch of electronics, spintronics, the spin of electrons plays the central role. There are several reasons for intense activity in this emerging field, promising better performance with smaller power consumption\cite{Wolf2001,Zutic2004,Rashba2007}. Qubits represented with ordinary spin of electrons seems to be a natural choice\cite{Awschalom2013} and gated semiconductor devices based on quantum dots and quantum wires configurations are believed to be one of the most suitable candidates for the realisation of quantum computers. However, the main challenge  is the possibility to manipulate spin of a single electron in the absence of magnetic fields, which can not be applied locally in a small region.

The application of devices using the spin-orbit interaction is a possible solution of this problem in semiconductor heterostructures\cite{Zutic2004,Winkler2003,Engel2007}. There are two types of spin-orbit interaction present in such heterostructures, the Dresselhaus type\cite{Dresselhaus1955} of interaction which emerges due to bulk inversion asymmetry of a crystal, and the Rashba type spin-orbit interaction\cite{Rashba1960}  which is a consequence of structural inversion asymmetry of confining potential of the two-dimensional electron gas. The strength of the Rashba interaction can be tuned externally using voltage gates\cite{Nitta1997,Schapers1998} which makes this type of interaction suitable for spintronic devices. Most of proposals for spintronic devices are based on a diverse range of two-dimensional  semiconductor structures with spin-orbit,  electron-
electron Coulomb, electron-phonon or spin-spin interaction\cite{Wolf2001,Zutic2004,Nitta1999,Schliemann2003,Wunderlich2010,Stajic2013}.

Recently a possible solution to the spin qubit manipulation was developed for a linear quantum wire with time dependent spin-orbit interaction and driven by external time-dependent potential\cite{Cadez2013}. By the virtue of the exact solution one can construct analytically the corresponding non-adiabatic non-Abelian Anandan phase\cite{Anandan1988} which opens the possibility of holonomic qubit transformations. Analytical solutions enable exact analysis of arbitrary fast driving in one dimension with tunable Rashba coupling -- but due to the fixed axis of rotation, spin transformations on the Bloch sphere are severely restricted\cite{Cadez2014}. 

Limitations posed by fixed axis of spin rotation in linear systems can be eliminated in quantum ring structures. These systems are still geometrically simple, but exhibit a rich range of phenomena, for example a well known Aharonov-Bohm effect and persistent currents in magnetic fields\cite{Buttiker1983,Fuhrer2001}, the Aharonov-Casher effect in the presence of the spin-orbit interaction\cite{Aronov1993,Qian1994,Malshukov1999,Richter2012} or the spin geometric phase manipulation\cite{Nagasawa2013,Saarikoski2015}. Recently an exact solution for the time-dependent wavefunction propagating around a quantum ring has been derived\cite{Kregar2015}. Proposed arbitrary transformations on Bloch sphere are achieved by a segment propagation of Kramers-doublet qubits using external driving potential and time-dependent Rashba coupling.
Although the first correct derivation of the Hamiltonian for electrons on a ring in the presence of the Rashba interaction has been derived already a decade ago\cite{Meijer2002}, the analysis of these systems still represents an interesting challenge for analytical approaches. 
  
In this paper we present the derivation of an exact unitary transformation which maps the problem of a ring in the presence of the Rashba spin-orbit interaction and in an external magnetic field to a simple bare ring without spin-orbit terms. The transformation is composed of three steps and we show how these are related to the symmetries of the ring. As an example of the application of the transformation we present exact results for energy levels and expectation value of spin for an electron on the ring. It is also shown how the transformation can be applied for ring systems with additional external electric potentials\cite{Kregar2015}.

\section{Symmetry properties}
Consider two-dimensional electron gas, described by the Hamiltonian with the Rashba spin-orbit coupling $\alpha_R$ and in the presence of magnetic field $\vec{B}$, vector potential $\vec{A}$ and external confining potential $V_e(\vec{r}\,)$,
\begin{equation}
\label{eq:1_1_1}
H = \frac{1}{2 m} \left(\vec{p} - e \vec{A} \right)^2 + V_e( \vec{r}\,)+\alpha_R \vec{\sigma} \cdot \vec{e}_z \times \left(\vec{p} - e \vec{A} \right) +  \frac{\mu_B g}{2} \vec{\sigma} \cdot \vec{B},
\end{equation}
with  the effective electron mass $m$, charge $e$, standard Pauli spin operator $\vec{\sigma}$, the Land\' e factor $g$ and Bohr magneton $\mu_B$. This Hamiltonian can be simplified for the use on a narrow ring of radius $R$, described by the polar angle $\varphi$, leading to the ring Hamiltonian\cite{Meijer2002}
\begin{equation}
\label{eq:Hringdim}
H =\epsilon \left( i \partial_\varphi + \phi_m \right)^2  -\epsilon \alpha \left[ \sigma_\rho(\varphi) \left(i \partial_\varphi +\phi_m\right) + \frac{i}{2} \sigma_\varphi(\varphi) \right] + b \sigma_z +V(\varphi),
\end{equation}
with redefined parameters
\begin{equation}
\label{eq:def1}
\epsilon \equiv \frac{\hbar^2}{2 m R^2}, \qquad \alpha \equiv \frac{2 m R \alpha_{R}}{\hbar},\qquad b \equiv \frac{\mu_B g B}{2}, \qquad  \phi_m \equiv \frac{\phi}{\phi_0},
\end{equation}
where $\phi_0 = \hbar/e$ is the flux quantum and magnetic flux through the ring $\phi = \oint \vec{A} \cdot d \vec{r}$ is calculated as a contour integral of the vector potential around the ring. Effective potential $V(\varphi)$ is obtained by mapping the external potential $V_e(\vec{r}\,)$, defined in two dimensions, to the ring.

Spin interaction is expressed by transformed Pauli matrices $\sigma_\rho$ and $\sigma_\varphi$, which are an angle-dependent linear combination of ordinary Pauli matrices $\sigma_x$ and $\sigma_y$,
 \begin{align}
\label{eq:1_1_3}
\sigma_\rho\left( \varphi \right) &=\phantom{-} \sigma_x \cos \varphi + \sigma_y \sin \varphi,
\nonumber \\
\sigma_\varphi\left( \varphi \right) &=  -\sigma_x \sin \varphi + \sigma_y \cos \varphi.
\end{align}

In the absence of the potential $V(\varphi)$, the Hamiltonian Eq.~(\ref{eq:Hringdim}) exhibits interesting symmetry properties. Since the electric field, generating the Rashba coupling, is perpendicular to the ring plane, the Hamiltonian is invariant with respect to the rotation around the $z$-axis. It is important to note that this rotation consists of two terms. A rotation in real space by angle $\varphi_0$ is expressed as a shift of the coordinate $\varphi$ by the shift operator $T^\dagger(\varphi_0) = \exp(- \varphi_0 \partial_\varphi)$. However, the Hamiltonian is not invariant under this transformation,
\begin{equation}
T^\dagger(\varphi_0) H(\varphi) T(\varphi_0) = H(\varphi - \varphi_0) \neq H(\varphi).
\end{equation}
To construct a complete shift transformation, also the spin rotation around the $z$-axis should be added,  $U_z^\dagger(\varphi_0) = \exp(-\frac{\varphi_0}{2} \sigma_z)$. Combined transformation with both operators $T^\dagger_{rot}(\varphi_0)= T^\dagger(\varphi_0)U_z^\dagger(\varphi_0)$ transforms the Hamiltonian to its original form,
\begin{equation}
T^\dagger_{rot}(\varphi_0) H(\varphi) T_{rot}(\varphi_0) = H(\varphi).
\end{equation}
We notice that this combined rotation is generated by the $z$-component of the total angular momentum
\begin{equation}
T^\dagger_{rot}(\varphi_0) = \exp\left(-i \frac{\varphi_0 J_z}{2} \right),
\end{equation}
where the total angular momentum and the $z$-component are given by
\begin{align}
\vec{J} &= \vec{L} + \vec{s} = -i \hbar \partial_\varphi \vec{e}_z + \frac{\hbar}{2} \vec{\sigma}, \nonumber \\
J_z &= L_z + s_z = -i \hbar \partial_\varphi + \frac{\hbar}{2} \sigma_z.
\end{align}
The symmetry can therefore also be manifested by the commutator of the Hamiltonian and $J_z$
\begin{equation}
\left[ J_z, H \right] = 0.
\end{equation}
Additionally the Hamiltonian commutes also with the square of the total angular momentum $\vec{J}$,
\begin{align}
J^2 = \vec{J} \cdot \vec{J} &= \hbar^2 \left( -\partial^2_\varphi - i \sigma_z \partial_\varphi + \frac{3}{4} \right), \nonumber \\
\left[ J^2, H \right] &=0.
\end{align}

\section{Unitary transformations}

Seeking the solution of stationary Schr\"odinger equation corresponding to the Hamiltonian Eq.~(\ref{eq:Hringdim})  is significantly simplified by applying an appropriate unitary transformation to the Hamiltonian. The original equation
\begin{equation}
\label{eq:2_4_stat_sch}
H \psi(\varphi) = E \psi(\varphi),
\end{equation}
thus should be transformed to the equation
\begin{equation}
H' \psi'(\varphi) = E \psi'(\varphi)
\end{equation}
with the same energy, but with different Hamiltonian and eigenstates
\begin{equation}
\label{eq:1_2_21}
H' = U H U^\dagger, \quad \psi'(\varphi) = U \psi(\varphi).
\end{equation} 

The transformation consists of three steps. First we use the well known Peierls transformation $U^\dagger_\phi = \exp( i \phi_m \varphi)$ to eliminate the magnetic flux
\begin{equation}
U_\phi H U^\dagger_\phi = - \epsilon \partial_\varphi ^2  -i \epsilon \alpha \left[ \sigma_\rho(\varphi)  \partial_\varphi+ \frac{1}{2} \sigma_\varphi(\varphi) \right] + b \sigma_z + V(\varphi).
\end{equation}
Note that this transformation affects only the terms containing $\partial_\varphi$.

For the next step, we notice that rotated Pauli matrices $\sigma_\rho$ and $\sigma_\varphi$ can be expressed with the use of the spin rotation $U_z^\dagger$
\begin{align}
\label{eq:1_1_38}
\sigma_{\rho}(\varphi) &= U_z^\dagger(\varphi) \sigma_{x} U_z(\varphi),  \nonumber \\
\sigma_{\varphi}(\varphi) &= U_z^\dagger(\varphi) \sigma_{y} U_z(\varphi).
\end{align} 
The inverse transformation will therefore eliminate the $\varphi$-angle dependence from spin terms in the Hamiltonian
\begin{align}
U_z U_\phi H U^\dagger_\phi U_z^\dagger =& \epsilon\left( i \partial_\varphi + \frac{1}{2} \sigma_z \right)^2  - i \epsilon \alpha  \sigma_x  \partial_\varphi + b \sigma_z + V(\varphi)= \nonumber \\
 =& \epsilon \left( i \partial_\varphi \right)^2  + i \partial_\varphi \left( -\alpha \sigma_x + \epsilon \sigma_z \right) + b \sigma_z + \frac{\epsilon}{4} + V(\varphi).
\end{align}
In the derivation of this last result the following derivatives of rotated Pauli matrices were applied
\begin{equation}
\label{eq:1_1_42}
 \partial_\varphi   \sigma_\rho = \sigma_\varphi, \quad  \partial_\varphi   \sigma_\varphi = -\sigma_\rho.
\end{equation}
The Zeeman term $ b \sigma_z$ remains unaffected due to the commutation of $\sigma_z$ and $U_z^\dagger$, as well as the potential term $V(\varphi)$, which is assumed spin independent.

\begin{figure}[hbt]
\centerline{\psfig{file=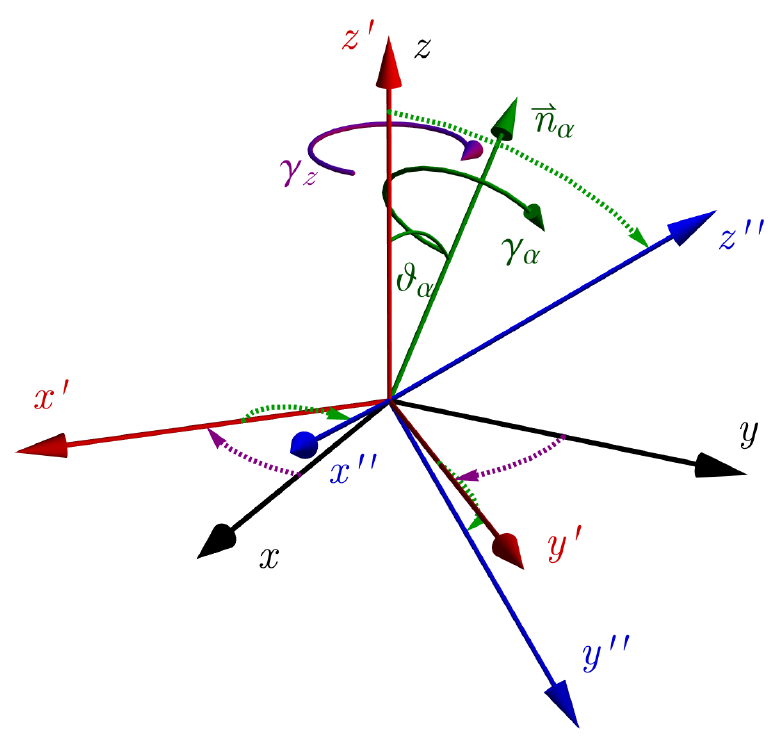,scale=1}}

\vspace*{8pt}
\caption{Schematic representation of transformations $U_z$ and $U_\alpha$ as spin rotations. $U_z$ (purple) rotates original spin frame (black) to a rotated frame (red), which is further rotated by $U_\alpha$ (green) to the final frame (blue).}
\label{fig:spin_rotation}
\end{figure}

The spin-orbit term of the Hamiltonian is now independent of $\varphi$ and can therefore be eliminated by a spin rotation around a tilted axis\cite{Lobos2008}
\begin{equation}
U_\alpha^\dagger = \exp \left(i \frac{\varphi}{2} \vec{\alpha} \cdot \vec{\sigma} \right), \quad \vec{\alpha} = \left( -\alpha,0,1\right). 
\end{equation}
Emerged Hamiltonian, fully transformed using $U = U_\alpha U_z U_\phi$, is then given by
\begin{equation}
H' =U H U^\dagger =  - \epsilon \partial_\varphi ^2  - \frac{1}{4}\epsilon \alpha^2 + V(\varphi)+ b U_\alpha \sigma_z U_\alpha^\dagger.
\end{equation}
Transformations $U_z$ and $U_\alpha$ are depicted as spin rotations in Fig.~\ref{fig:spin_rotation}. $U_z$ rotates spin around the $z$-axis (red) for an angle $\gamma_z = -\varphi$ (purple) while $U_\alpha$ rotates around the direction $\vec{\alpha}$ by angle $\gamma_\alpha = \varphi \sqrt{1+\alpha^2}$ (green). In the limiting case $\alpha = 0$ the rotation axes coincide and rotations cancel out.

In the absence of the Zeeman term, which now is $\varphi$-dependent, the Hamiltonian has a very simple form with kinetic energy, angle-dependent potential and a constant term. This means that if we are able to find the eigenstates $\psi'(\varphi)$ of a one-dimensional system with particular external potential in the absence of the Rashba coupling, the solution can be mapped to an eigenstate of the original Rashba ring,
\begin{equation}
\label{eq:psi_trans}
\psi(\varphi) = U^\dagger \psi'(\varphi).
\end{equation}

\section{Modified boundary conditions}

However, the eigenstates are determined not only by the Hamiltonian, but also by the boundary conditions. For the eigenstates of the Rashba ring Hamiltonian, periodic boundary conditions apply
\begin{equation}
\psi(\varphi+2 \pi) = \psi(\varphi).
\end{equation}
Using Eq.~\eqref{eq:psi_trans}, it is clear that the same does not hold for an eigenstate $\psi'(\varphi)$ of the transformed Hamiltonian. The transformation $U^\dagger$ must be taken into account, leading to
\begin{equation}
U^\dagger(0) \psi'(0) = U^\dagger(2\pi) \psi'(2\pi).
\end{equation}
This expression can be simplified further using $U^\dagger(0) = 1$ and $U_z^\dagger(2\pi) = -1$,
\begin{equation}
\psi'(0) = - e^{2\pi i \phi_m } U_\alpha^\dagger(2 \pi) \psi'(2 \pi).
\end{equation}
Since $U_\alpha^\dagger(2 \pi)$ is a spin operator, this boundary conditions will lead to non-trivial spin properties of the eigenstates.

\section{Eigenstates of free electrons: energy and spin properties}

As an example of the application of the formalism, we solve the Schr\"odinger equation for a ring without external potential and the Zeeman term. The transformed Hamiltonian is in this case given by
\begin{equation}
H' =- \epsilon \partial_\varphi ^2  - \frac{1}{4}\epsilon \alpha^2.
\end{equation}
The eigenstates have a form of plane waves with constant spinor 
\begin{equation}
\psi'_{k s}(\varphi) = e^{i k \varphi} \chi^*_s.
\end{equation}
Both, the wave number $k$ and the spinor $\chi_s^*$, are determined by modified boundary conditions,
\begin{equation}
\label{eq:bound_cond1}
 - e^{2 i \pi \left( \phi_m + k\right) } U_\alpha^\dagger(2 \pi) \chi^*_s = \chi^*_s,
\end{equation}
which is essentially the eigenstate problem for the operator $U_\alpha^\dagger(\varphi)$
\begin{equation}
U_\alpha^\dagger(\varphi) \chi^*_s = \lambda_s \chi_s^*.
\end{equation}
Since $U_\alpha^\dagger$ is a spin rotation operator, the eigenvalues are exponents of the rotation angle and eigenspinors point into the direction of the rotation axis
\begin{equation}
\lambda_s = e^{s \varphi \phi_\alpha}, \quad \chi_s^* = U_y^\dagger(\vartheta_\alpha) \chi_s, \quad \tan \vartheta_\alpha = -\alpha.
\end{equation}
Spin quantum number takes values $s = \pm 1/2$, corresponding to the basis spinor $\chi_s$, quantised in the $z$-direction with $1/2 \equiv \uparrow$ and $-1/2 \equiv \downarrow$, respectively. $\phi_\alpha = \sqrt{1 +\alpha^2}$ is the Aharonov-Casher flux\cite{Frustaglia2004,Nagasawa2012} and $U_y(\vartheta_\alpha) = \exp\left( -\frac{\vartheta_\alpha}{2} \sigma_y \right)$ is  spin rotation around the $y$-axis.
 
Applying this result to Eq.~(\ref{eq:bound_cond1}),
 \begin{equation}
  e^{2 i \pi \left( \phi_m + k + s \phi_\alpha  \right) } \chi^*_s = \chi^*_s,
 \end{equation}
it becomes clear that the wave vector $k$ also depends on $s$,
\begin{equation}
k_s = j - \phi_m - s \phi_\alpha, \quad j -1/2 \in \mathbb{Z}.
\end{equation}
Here we introduced new quantum number $j$ which takes half-integer values and corresponds to the angular momentum of the eigenstates. Transformed eigenstates are then
\begin{equation}
\psi'_{j s}(\varphi) = e^{i \left( j - \phi_m - s \phi_\alpha \right) \varphi} \chi_s^*,
\end{equation}
leading to original eigenstates
\begin{equation}
\psi_{j s}(\varphi) = U^\dagger(\varphi) \psi'_{j s}(\varphi) = e^{i j \varphi} U_z^\dagger(\varphi) U_y^\dagger(\vartheta_\alpha) \chi_s.
\end{equation}
Note that the term $\phi_m$  in the exponent is cancelled out with the transformation $U^\dagger_\phi$ and $s \phi_\alpha$ with $U_\alpha^\dagger$. If written by components, it is evident that this reproduces know results from\cite{Splettstoesser2003,Frustaglia2004,Sheng2006,Chen2007,Berche2010}. The corresponding eigenenergies are then
\begin{align}
\label{eq:1_2_45}
E_{j s} =& \epsilon \left(k_s^2 - \frac{\alpha^2}{4} \right)  = \nonumber \\
= & \epsilon \left(j -  \phi_m \right)^2 - 2 \epsilon s \left(j -  \phi_m \right) \phi_\alpha + \frac{\epsilon}{4}.
\end{align}

\begin{figure}[hbt]
\centerline{\psfig{file=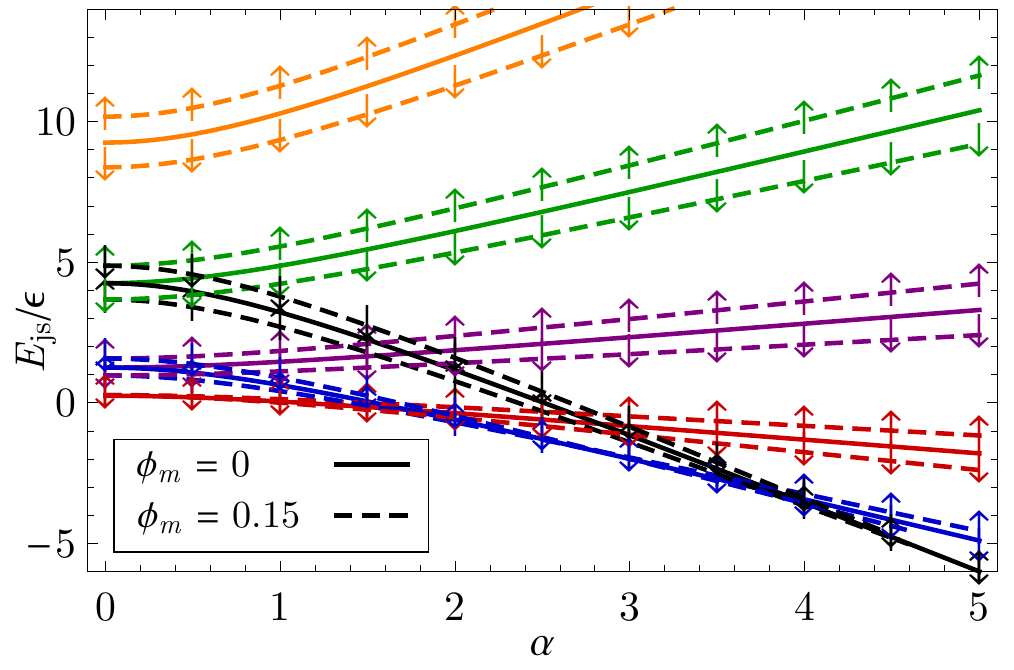,scale=0.8}}
\vspace*{8pt}
\caption{Eigenenergies $E_{j s}$ of the Kramers doublets as a function of $\alpha$: $(j,s) = \pm\left(\frac{1}{2}, \frac{1}{2} \right)$ in red, $\pm\left(\frac{1}{2}, -\frac{1}{2} \right)$ in purple, $\pm\left(\frac{3}{2}, \frac{1}{2} \right)$ in blue, $\pm\left(\frac{3}{2}, -\frac{1}{2} \right)$ in green,  $\pm\left(\frac{5}{2}, \frac{1}{2} \right)$ in black and $\pm\left(\frac{5}{2}, -\frac{1}{2} \right)$ in orange colour. In absence of magnetic flux (solid line), the energies are degenerate, while in presence of flux the energy of spin $s=\pm 1/2$ is split (dashed line). Spin of split energy levels is denoted by arrows $\uparrow=1/2$ and $\downarrow  =-1/2$.}
\label{fig:energy}
\end{figure}

In the absence of magnetic flux, states are symmetric to time reversal, resulting in Kramers degenerate pairs\cite{Kramers1930}. In our case, these are states with opposite sign of quantum numbers $j$ and $s$, $E_{j, s} = E_{-j, -s}$. These states are depicted as solid lines in Fig.~\ref{fig:energy}, where energy of states is plotted as a function of the Rashba coupling. When magnetic flux is added, time-reversal symmetry breaks and degeneracy is lifted (dashed lines in same figure).

It is easy to verify that the eigenstates are also eigenstates of $J_z$ with eigenvalues $j \hbar$,
\begin{align}
J_z \psi_{j s}(\varphi) &= \hbar \left( - i \partial_\varphi + \frac{1}{2} \sigma_z\right)  e^{i j \varphi}  U_z^\dagger(\varphi) U_y^\dagger \chi_s = \nonumber \\
&=\hbar \left( \left[j -  \frac{1}{2} \sigma_z \right] + \frac{1}{2} \sigma_z\right)  e^{i j \varphi}  U_z^\dagger(\varphi) U_y^\dagger \chi_s= \nonumber \\
&=j \hbar\psi_{j s}(\varphi).
\end{align}
For the sake of completeness we also calculate expectation values of the operator $J^2$, for which the eigenstates are $\psi_{j s}$,
\begin{equation}
J^2 \psi_{j s}(\varphi) =  \left( j^2 + \frac{1}{2} \right)\hbar^2 \psi_{j s}(\varphi).
\end{equation}

\begin{figure}[hbt]
\centerline{\psfig{file=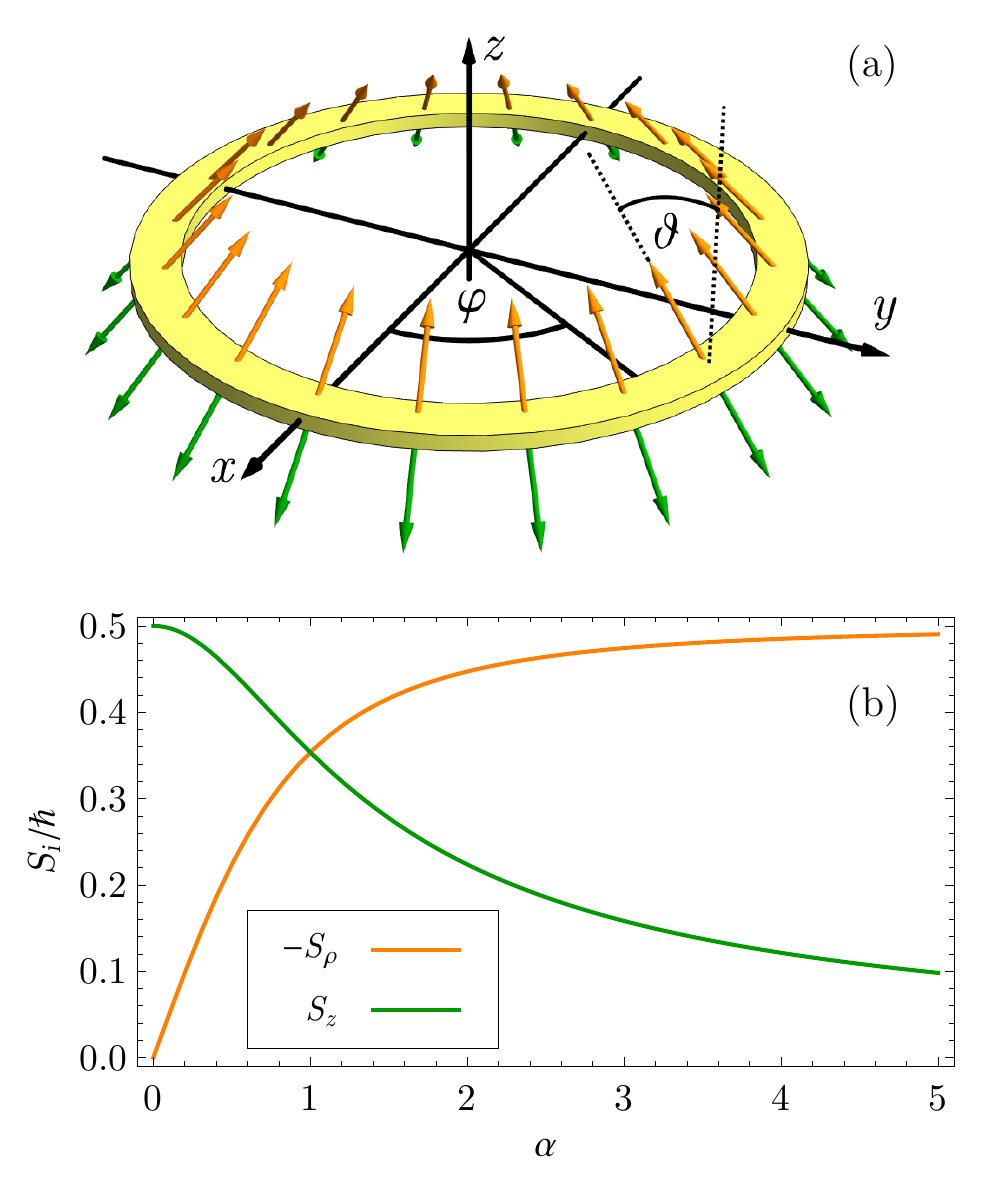,scale=0.8}}
\vspace*{8pt}
\caption{(a) Schematic representation of local expectation values  of spin $\vec{S}_{j s}$ on a ring with $\alpha = 0.5$. Orange arrows correspond to $s=1/2$ and green to $s=-1/2$. (b) Radial and $z$-direction component of local expectation values of spin as a function of the Rashba coupling for $s=1/2$. The values for $s=-1/2$ have opposite sign.}
\label{fig:spin}
\end{figure}

Due to the structure of spin operators $U_z^\dagger$ and $U_y^\dagger$ the eigenstates exhibit some interesting spin properties. These are clearly demonstrated if local spin expectation value is calculated, 
\begin{equation}
\vec{S}_{j s}(\varphi) = \psi_{j s}^\dagger(\varphi) \frac{\hbar}{2} \vec{\sigma} \,\psi_{j s}(\varphi).
\end{equation}
This quantity represents the expectation value of the spin of the system, if the electron is strongly confined around the position $\varphi$. By applying unitary transformations the local spin expectation value is not difficult to calculate,
\begin{eqnarray}
\vec{S}_{j s}(\varphi)& =& \chi_s^\dagger U_y(\vartheta_\alpha) U_z(\varphi) \frac{\hbar}{2} \vec{\sigma} \,U_z^\dagger(\varphi) U_y^\dagger(\vartheta_\alpha)\chi_s= \nonumber\\
&=&s \hbar \left[\cos (\varphi ) \sin(\vartheta _{\alpha }), \sin (\varphi ) \sin(\vartheta _{\alpha }), \cos \left(\vartheta _{\alpha }\right)\right].
\end{eqnarray}
The spin forms a crown-like structure, shown in Fig.~\ref{fig:spin}(a). Due to the Rashba coupling, the spins are tilted from the $z$-direction towards the centre of the ring for $s=1/2$ and from $-z$ away from the centre for $s=-1/2$, respectively. Radial and $z$-component of local spin do not depend on angular momentum $j$, but only on the Rashba coupling, as depicted in Fig.~\ref{fig:spin}(b).

Ordinary expectation values of spin are calculated as an integral of $\vec{S}_{j s}(\varphi)$ over the coordinate $\varphi$. $x$ and $y$-components, proportional to sine and cosine of $\varphi$, vanish in the process and only the $z$-component remains finite
\begin{equation}
\left\langle \vec{S}_{j s} \right\rangle = \frac{1}{2 \pi}\int_{0}^{2\pi} \vec{S}_{j s}(\varphi) d \varphi =s \hbar \left[ 0, 0, \cos( \vartheta_\alpha) \right].
\end{equation}

\section{Summary}

We have presented a simple formalism for the analysis of electron properties on rings with spin-orbit interaction and in the presence of external magnetic or electric fields. The essence of the formalism is in the application of a unitary transformation, composed of three separate steps, where it is essential to introduce the total angular momentum and to incorporate the boundary conditions consistently. 
As an example, explicit expressions for the energy and spin of eigenstates of an electron on a ring in the absence of electric fields are shown, in agreement with known results. It is then straightforward to apply the formalism -- the unitary transformation  -- also to systems where the electron is confined in additional external, possibly time-dependent, potentials. After such a transformation the Rashba coupling is incorporated into the transformation operators and eliminated from the model, which can significantly simplify the formalism\cite{Kregar2015}.

\section*{Acknowledgements}

The authors thank J. H. Jefferson for discussions and acknowledge support from the Slovenian Research Agency under contract no. P1-0044.

\end{document}